\documentclass[aps,groupedaddress,showpacs]{revtex4}
\usepackage{amsmath}
\usepackage{graphics}
\usepackage{hyperref}

\begin{document}
\bibliographystyle{revtex}
\preprint{TP-CB-399}

\title{Pipeline model of a Fermi--sea electron pump}

\author{Fernando Sols}
\affiliation{Dpto.\ de F\'{\i}sica Te\'orica de la Materia Condensada,
             C-V and Instituto de Ciencia de Materiales ``Nicol\'as Cabrera'', 
	 Universidad Aut\'onoma de Madrid, E-28049 Madrid, Spain}
\author{Mathias Wagner}
\email[]{wagner@phy.cam.ac.uk}
\affiliation{Hitachi Cambridge Laboratory, Madingley Road,
             Cambridge CB3 0HE, United Kingdom}

\date{\today}

\begin{abstract}
    The use of a band offset between the two leads of an electron pump driven 
    by a local oscillating voltage is shown to increase the pump current 
    dramatically. The structure of the electron transmission suggests the 
    existence of dominant inelastic channels which we call 
    pipelines. This permits the formulation of a simple model that gives a 
    physical account of the numerical results for a realistic device. A 
    spectral analysis reveals the pump current to be carried by  
    scattering states with initial energy deep within the Fermi sea and not 
    at its surface, thereby rendering the effect insensitive to 
    temperature. We show this is compatible with the current flowing
    near the Fermi surface in the leads.
\end{abstract}

\maketitle

\bibliographystyle{plain}

\section{Introduction}

An electron pump stimulates directed electron motion 
by locally acting on the electron system. This current 
rectification is achieved by combining nonlinear ac driving with either 
absence of inversion symmetry in the device, or lack of time-reversal 
symmetry in the ac signal. The range of possible electron pumps 
includes the Thouless pump \cite{swit99,thou83} based on Archimedes' 
water pump, driven quantum-dots \cite{staf96}, electron turnstiles 
\cite{geer90}, and the recently proposed Fermi-sea pump \cite{Wagner99a}, 
which makes essential use of a band offset between the two leads. An 
electron pump may be viewed as a localized version of a ratchet 
\cite{magn93,bart94,zapa96} that has 
been stripped of its spatial periodicity, leaving nonlinearity and asymmetry 
as the only ingredients responsible for rectified electron motion. Here the 
role of quantum dissipation \cite{reim97} is played by the lack of 
phase coherence between the incoming scattering channels in the left and 
right leads.

\begin{figure}[t]
\centerline{\resizebox{11cm}{!}{\includegraphics{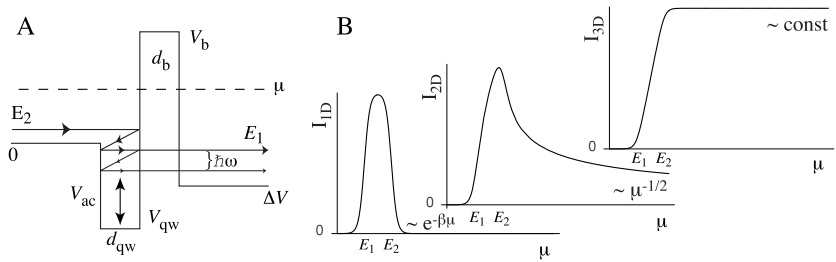}}}
\caption{A: Schematic potential profile for a Fermi-sea pump. The chemical
         potential $\mu$ is the same in both contacts. B: Pump currents in
	 1, 2, and 3 spatial dimensions as a function of the chemical
	 potential $\mu$ for fixed driving as determined by a single--pipeline
	 model.}
\label{Structure}
\end{figure}

In Ref.~\cite{Wagner99a}, we studied an electron pump whose schematic band 
diagram is shown in panel A of 
Fig.~\ref{Structure}. In this model, a quantum well is driven harmonically by 
an external ac potential $V_{\rm ac}\cos\omega t$ provided, for instance, by 
some external gates. Adjacent to the well is a static barrier, and the 
overall potential profile features a band offset $\Delta V$ between 
the left and right leads. One should note that this band offset is {\it not}
an external bias but rather is due to, for instance, different materials
used. The chemical potentials in the contacts to the left and
right are taken to be the same. Any dc current flowing is thus due to the
effect of the driving ac force alone. It was noted by us \cite{Wagner99a} 
that the spatial asymmetry of the model, in particular the band offset 
$\Delta V$, is vital for a large pump current to exist --- boosting 
it by as much as three orders of magnitude.
Crucially, we found the pump current to be carried by electrons with their  
initial energy narrowly distributed over a fixed interval approximately 
$\hbar\omega$ wide which, for sufficiently high Fermi energies, may stay well 
below the Fermi surface, thereby rendering the total current insensitive to 
temperature. 
The physical explanation for this peculiar behavior lies in the structure of 
{\it pipelines} displayed by the total transmission probability of the device 
\cite{Wagner99a}. Pipelines are 
pairs of left and right scattering channels, of dissimilar energies 
$E_1 \neq E_2$, that are strongly coupled. Thus, an electron coming, say, from 
the left with energy $E_2$ has a relatively high probability of being 
transmitted to the right with energy $E_1 < E_2$ (see Fig.~\ref{Structure}). 
In the most important cases, 
$E_2= E_1+ \hbar\omega$. Due to time-reversal symmetry, $V_{\rm ac}(z,t)$ = 
$V_{\rm ac}(z,-t)$, these pipelines obey microreversibility:
The probability for an electron to go from 
$E_2$ to $E_1$ can be proved to be the same as for the reverse 
process, i.e., $T_{\rightarrow}(E_1,E_2)$ $\equiv$ 
$T_{\leftarrow}(E_2,E_1)$.

We further showed that a good deal of the device physics 
can be explained in terms of the pipeline structure. In particular, one 
may assume the existence of a single pipeline and reproduce analytically 
many of the transport features obtained from a sophisticated numerical 
calculation. 
The purpose of this paper is to explore this pipeline model in further detail 
by performing an analytical study that leads to a deeper understanding 
of the physics involved. In particular, we analyze the different 
ways in which the total pump current can be decomposed spectrally.  
As in Ref.~\cite{Wagner99a}, we will discuss the properties of the model 
shown in Fig.~\ref{Structure}
in 1, 2 and 3 spatial dimensions, assuming separability of the 
Hamiltonian $H(t) = -(\hbar^2/2m) \Delta + V_0(z) + V_{\rm ac}(z,t)$.

To evaluate the electronic current, we need to determine the scattering
states in the presence of the driving field. Broadly, these states fall into two
classes: Scattering states with longitudinal kinetic energy $E_z$ in the 
incident 
channel much larger than the photon energy $\hbar\omega$ of 
the driving field are far away from the conduction band edge, and hence will 
be insensitive to fine details such as band offset or the presence of 
a shallow quantum well. 
Consequently, these states will not contribute to a net pump current, which is 
in
agreement with a recent analysis of pumps in this regime \cite{Pedersen98a}.
This is the more commonly studied regime. However, scattering states with 
{\it small} longitudinal kinetic energies of the order of $\hbar\omega$ are very 
sensitive to changes in the band profile and, in particular, have a high 
probability of the electron getting trapped in the quantum well. The underlying 
mechanism is that an 
electron incident, say, from the left with an energy $E_2$ less than 
$\hbar\omega$ 
may, upon reflection at the barrier, emit a photon and thus lose sufficient 
energy to get caught in the quantum well. Unless it subsequently manages to 
absorb another photon, 
the only chance to escape from the quantum well is by tunneling to the right at 
energy $E_1 = E_2-\hbar\omega$. This is the mechanism underlying the structure 
of 
asymmetric pipelines depicted in panel A of Fig.~\ref{Structure}.

In Sect.~\ref{Sec:The-pipeline-model} we present the pipeline model and derive 
some 
of its analytical properties, deferring the spectral properties of the current
within this model to Sect.~\ref{Sec:Spectral-analysis}. In 
Sect.~\ref{Sec:Numerical-results} we discuss numerical results for a 
realistic device and compare them with the analytical predictions of 
Sects.~\ref{Sec:The-pipeline-model} and \ref{Sec:Spectral-analysis}. 
Section \ref{Sec:Conclusions} contains a summary and some conclusions.

\section{The pipeline model}
\label{Sec:The-pipeline-model}

Consider a single pipeline of strength $T_p$ connecting the energies 
$E_1$ on the right and $E_2$ to the left. Assuming incident 
electrons approaching the device outside these two channels to be reflected with 
unit probability, we write for the scattering probabilities
\begin{eqnarray}
T_{\rm LR}(E_z,E'_z) &=& 
	   T_p \delta(E_z-E_2) \delta(E'_z-E_1) \label{TLR} \\
T_{\rm RL}(E_z,E'_z) &=& 
	   T_p \delta(E_z-E_1) \delta(E'_z-E_2) \label{TRL} \\
R_{\rm LL}(E_z,E'_z) &=& \delta(E_z-E'_z) -
	   T_p \delta(E_z-E_2) \delta(E'_z-E_2) \label{RLL} \\ 
R_{\rm RR}(E_z,E'_z) &=& \delta(E_z-E'_z) -
	   T_p \delta(E_z-E_1) \delta(E'_z-E_1). \label{RRR}
\end{eqnarray}
The model (\ref{TLR})-(\ref{RRR}) satisfies all the required symmetry 
properties, including unitarity. If, for a given energy $E_z$, we define
$T_{\rm net} \equiv T_{\rm \rightarrow} - T_{\rm \leftarrow}$ where
\begin{eqnarray}
T_{\rm \rightarrow}(E_z) = \int dE'_z T_{\rm LR}(E_z,E'_z) \quad,\quad
T_{\rm \leftarrow}(E_z)  = \int dE'_z T_{\rm RL}(E_z,E'_z),
	\label{eq:Tnet}
\end{eqnarray}
then we have \cite{Wagner99a}
\begin{eqnarray}
	T_{\rm net}(E_z) = 
	   T_p \left[\delta(E_z-E_2) - \delta(E_z-E_1)\right] .
	\label{eq:PipelineModel}
\end{eqnarray}
A special property of our pump is that the ac driving is localized in
space. 
Consequently, the distribution of {\it incident} electrons
is not affected by the driving and can be taken to be in thermal
equilibrium with the contact reservoirs. Furthermore, if we assume
the temperature of the two contacts to be the same, and that no
dc bias is applied, the distribution of incident electrons will be 
the same on both sides and given by a Fermi function $f(E-\mu)$, where
$\mu$ is the chemical potential in the contacts. With this the total
current can simply be written as \cite{Wagner99a}
\begin{eqnarray}
	I &=& \int_{\Delta V}^\infty dE f(E-\mu) J(E)
	\label{eq:Current}  \\
J(E) &=& \frac{2e}{h} \int_{\Delta V}^{E} dE_z D_{\bot}(E-E_z) T_{\rm net}(E_z) 
.
	\label{Spectral}
\end{eqnarray}
where $D_{\bot}$ is the density of states in the dimensions 
perpendicular to the direction of transport. In the single-pipeline model, 
Eqs.~(\ref{eq:PipelineModel})--(\ref{Spectral})  
yield
\begin{equation}
I=\frac{2e}{h}T_p \int_{0}^{\infty} dE_{\bot} D_{\bot}(E_{\bot}) 
[f(E_{\bot}+E_2-\mu)-f(E_{\bot}+E_1-\mu)] .
\label{current-pipeline}
\end{equation}

For one spatial dimension, Eq.~(\ref{current-pipeline}) translates into
\begin{eqnarray}
	I_{\rm 1D} = \frac{2e}{h} T_p 
	      \left[f(E_2-\mu) - f(E_1-\mu)\right] , 
	\label{eq:Current-1D}
\end{eqnarray}
which has a peak at $\mu$ = $(E_1+E_2)/2$, and an {\it exponential} 
decay for $\mu$ $\gg$ $kT$ as shown in panel B of Fig.~\ref{Structure}.
In 2D we find
\begin{eqnarray}
	I_{\rm 2D} \approx {2 e \over h^2} T_p \sqrt{2 \pi m \beta}
	     \Bigl[
	         {\rm Li}_{-{1\over 2}}(-{\rm e}^{\beta\mu})
		        (E_2-E_1) 
          -{\beta\over 2} {\rm Li}_{-{3\over 2}}(-{\rm e}^{\beta\mu})
		        (E_2^2-E_1^2)
	      \Bigr] ,
	\label{eq:Current-2D}
\end{eqnarray}
where Li is the polylogarithm function \cite{Lewin81a}. Expanding for 
$\mu$ $\gg$ $kT$ we find that in 2D the pump current decays only 
{\it algebraically} as $1/\sqrt{\mu}$.
Finally, in 3D we obtain
\begin{eqnarray}
	I_{\rm 3D} \approx {4\pi m e \over h^3} T_p 
	      \left[f(-\mu)(E_1-E_2) 
	  + {f'(-\mu)\over 2} (E_1^2-E_2^2)\right] .
	\label{eq:Current-3D}
\end{eqnarray}
For $\mu$ $\gg$ $kT$ one has $f(-\mu)$ $\approx$ 1 and $f'(-\mu)$ 
$\approx$ 0, i.e., the current in 3D becomes {\it independent} of $\mu$ in 
this limit, $I_{\rm 3D}$ = $-(4\pi m e/h^3) T_p (E_2-E_1)$, as seen in
panel B of Fig.~\ref{Structure}. Thus we find the non-trivial result that the 
pump
current does not necessarily decay with increasing chemical potential as one 
naively might expect.

\section{Spectral analysis}
\label{Sec:Spectral-analysis}

It is interesting to analyze the spectral function (\ref{Spectral}) that leads 
from Eq.~(\ref{eq:Current}) to Eq.~(\ref{current-pipeline}). Within the 
single--pipeline model, one obtains
\begin{equation}
J(E)=\frac{2e}{h} T_p [D_{\bot}(E-E_2)\theta(E-E_2)- D_{\bot}(E-E_1)\theta(E-
E_1)].
\label{spectral-pipeline}
\end{equation}
Eq.~(\ref{spectral-pipeline}) describes a function that is mainly localized in 
the pipeline region. In the particular case of three 
dimensions we have $D_{\bot}(E_{\bot})=2\pi m/h^2 \equiv D_0$ and 
Eq.~(\ref{spectral-pipeline}) yields a square function localized between $E_2$ 
and $E_1$. Combining this result with Eq.~(\ref{eq:Current}), which tells us 
that the total electric current $I$ is obtained by convoluting the spectral 
current density $J(E)$ with a thermal population of incoming electrons, we thus 
find that for $\mu \gg E_2$ the pump current is sustained by scattering states 
with incident energy well below the Fermi surface. An immediate consequence 
is that in this regime {\it the total pump current is insensitive to 
temperature}, even for $kT \sim \hbar\omega$. This remarkable result was noted 
in Ref.~\cite{Wagner99a} and will be explored further in the following within 
the framework of the analytical single--pipeline model.

First we note that it is possible to perform sensible spectral decompositions of 
the current different from (\ref{eq:Current})-(\ref{Spectral}). In 
Eq.~(\ref{eq:Current}), the integration variable $E$ refers to the {\it initial} 
electron energy. We may adopt an alternative viewpoint and write the current as 
an energy integral in which the variable $E$ denotes not the initial but the 
{\it actual} energy of the electrons flowing through a given region, say, in the 
left lead.

From the single--pipeline model (\ref{TLR})-(\ref{RRR}) we may write the 
pump current as
\begin{eqnarray}
I &=& \frac{2e}{h} \int dE_{\bot} \int dE_z \int dE'_z 
D_{\bot}(E_{\bot}) f(E_{\bot}+E_z-\mu) \nonumber \\
&& \times \, \left[\delta(E_z-E'_z) - R_{\rm LL}(E_z,E'_z) - T_{\rm 
RL}(E_z,E'_z)\right] 
\label{detailed}\\
&=& \frac{2e}{h} \int dE_{\bot} \int dE_z \int dE'_z 
D_{\bot}(E_{\bot}) f(E_{\bot}+E_z-\mu)
\nonumber \\
&& \times \, T_p \left[ \delta(E_z-E_2) - \delta(E_z-E_1)\right ] \delta(E'_z-
E_2).
\label{specific}
\end{eqnarray}
We wish to analyze the total current in terms of the final energy $ 
E=E_{\bot}+E'_z$ 
regardless of the initial longitudinal energy $E_z$ [the term  $\delta(E_z-
E'_z)$ in 
Eq.~(\ref{detailed}) describing the contribution from the stream of incoming 
electrons 
is explicitly cancelled by an identical term in Eq.~(\ref{RLL})]. With this goal 
in mind, 
we rewrite Eq.~(\ref{specific}) as
\begin{eqnarray}
I&=&\frac{2e}{h} T_p \int_0^{\infty} dE \int_0^{E} dE_z 
D_{\bot}(E-E_z) \nonumber \\
&&\times \, \left[f(E-E_z+E_2-\mu)- f(E-E_z+E_1-\mu)\right] \delta(E_z-E_2)
\end{eqnarray}
to obtain $I = \int_0^{\infty} J_L(E) \, dE$, where
\begin{eqnarray}
J_L(E) = \frac{2e}{h} T_p \theta(E-E_2) D_{\bot}(E-E_2)
\left[f(E-\mu)-f(E-\hbar \omega-\mu)\right].
\label{KLE}
\end{eqnarray}
In the 3D case, for which $D_{\bot}(E-E_2)=D_0$, it is clear that for $\mu \gg 
E_2$ the current spectrum in the left lead is essentially localized between 
$\mu$ and $\mu+\hbar\omega$, where we have taken $E_2-E_1=\hbar\omega$. A 
similar analysis for the current in the right lead yields
$I = \int_{\Delta V}^{\infty} J_R(E) \, dE$ with
\begin{eqnarray}
J_R(E) = \frac{2e}{h} T_p \theta(E-E_1) D_{\bot}(E-E_1)
\left[f(E+\hbar\omega-\mu)-f(E-\mu)\right],
\label{KRE}
\end{eqnarray}
which tells us that, except for thermal smearing, the net current is carried 
by electrons with energies between $\mu -\hbar\omega$ and $\mu$. We conclude 
that there is an average increase of $\hbar\omega$ in the energy of the 
electrons responsible for the current. This is expected, since $\hbar\omega$ 
is precisely the energy gained by electrons being transmitted from right to 
left. Thus we encounter the interesting result that, while the current is due 
to a lack of cancellation between scattering states with initial energies 
well below the Fermi level, the net current in the leads is carried by 
electrons near the Fermi surface. The energies of the independent electrons 
are modified in such a way that the current may have very different spectral 
properties depending on the way in which it is analyzed. The numerical
results discussed in the next section [see 
Fig.~\ref{AsymptoticSpectralCurrentDensity}] confirm these analytical 
predictions.

This result is exact but, perhaps, somewhat counterintuitive initially. One 
way of reconciling the two pictures is to try to understand them separately 
and to notice that they are, in fact, compatible. 

(i) If we view the net current as resulting from a lack of cancellation
between scattering states whose incident channel has total energy $E$, it is easy
to interpret Eqs.~(\ref{eq:Current}) and (\ref{spectral-pipeline}): If $E<E_1$ 
no state can benefit from the pipeline, and hence there is no contribution 
to the current at these energies. For $E_1<E<E_2$ only states coming from 
the right can take advantage of the pipeline and carry any current --- 
therefore, clearly, $J(E) \neq 0$ here. On the other hand, if $E>E_1,\,E_2$, 
there will be channels incident from {\it both} sides, whose energy in 
z-direction, $E_z$, matches the pipeline. Since the current from each side  
depends on the available density of states in the transverse direction, 
$D_{\bot}(E-E_z)$, and secondly, as one must fulfill the matching
conditions $E_z=E_2$ on the left 
and $E_z=E_1$ on the right, Eq.~(\ref{spectral-pipeline}) is readily 
understood. In this picture we thus have to conclude that $J(E)$ is localized 
in the pipeline region and furthermore, as gathered from Eq.~(\ref{eq:Current}), 
that $J(E)$ is independent of the (common) distribution function in the contacts.

(ii) In the alternative picture put forward in the present work we focus on, 
say, the current density $J_{\rm L}(E)$ on the left-hand side of the pump. 
Now $E$ refers to the local electron energy on that side, regardless of 
where the electron came from originally. It is clear that in this case the 
current results from a lack of cancellation between {\it left-going} electrons 
transmitted from the right-hand side via the pipeline, and therefore weighted 
with probability $f(E+E_1-E_2-\mu)$, and {\it right-going} transmitted electrons 
distributed according to $f(E-\mu)$. For a given energy $E>E_2$ the electrons 
contributing will be those that fulfill the pipeline-matching condition on the 
left-hand side, $E_z=E_2$, and their number will be proportional to the 
transverse density of states, $D_{\bot}(E-E_2)$. Hence the structure of 
Eq.~(\ref{KLE}), which yields a spectral function that is localized within 
$E_2-E_1$ = $\hbar\omega$ of the Fermi surface. Note that, in contrast to the 
first case, this definition of a spectral current density does explicitly 
depend on the distribution function.

Yet, there is no contradiction between pictures (i) and (ii). A net current 
results from a lack of cancellation between current-carrying states, but 
the natural way to identify these states depends on the way we group 
them into cancelling pairs. This can be done in more than one sensible way,
especially if the energy of the electron is not conserved. In (i) we compare 
{\it retarded scattering states} with initial energy $E$, whereas in (ii) we 
compare {\it scattering channels} of energy $E$ on a given side of the pump.
Both pictures are equally valid but, depending on the aspect one wishes to
explain, one may favor one over the other.

An interesting feature is that, while the total current is essentially 
independent of temperature for $\mu - E_2 \gg kT$, the local spectral 
distributions $J_L$ and $J_R$ are indeed sensitive to temperature. This 
suggests the existence of an effective sum rule that eliminates the 
temperature dependence of Eqs.~(\ref{KLE}) and (\ref{KRE}) when integrated 
over energy. It is easy to check this result explicitly in the 3D case, if 
one focuses on the limit $\omega \ll kT$. Then Eq.~(\ref{KRE}) can be 
approximated as
\begin{equation}
J_R(E) \approx (eT_pD_0/\pi) \, \omega f'(E+\hbar\omega/2-\mu),
\label{limit}
\end{equation}
which yields a result manifestly independent of temperature when integrated
over energy.

We end this section by noting two properties of the single--pipeline model. 
First, the left and right spectral functions can be shown to be exactly shifted 
by 
$\hbar \omega$,
\begin{equation}
J_L(E+\hbar\omega)=J_R(E),
\end{equation}
which is a straightforward consequence of (\ref{KLE}) and (\ref{KRE}). Second, 
there is a reflection symmetry around the chemical potential: If all energies 
are with reference to $\mu$,
\begin{equation}
J_L(E)=J_R(-E),
\label{symmetry}
\end{equation}
as long as $|E|<|E_2|$. This result is independent of the particular values of 
$E_1$ and $E_2$, and hence we expect it to survive in realistic situations
with a continuous distribution of pipelines and higher-order energy 
transfers of $E_2-E_1$ = 2$\hbar\omega$, 3$\hbar\omega$, \ldots.
This expectation is confirmed by the numerical results presented below
(see Fig.~\ref{AsymptoticSpectralCurrentDensity}).

\begin{figure}[t]
\centerline{\resizebox{12.0cm}{!}{\includegraphics{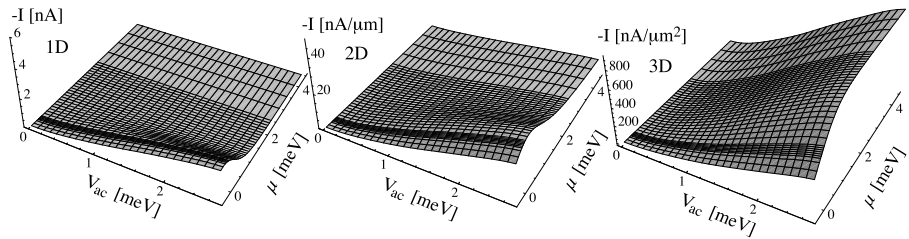}}}
\caption{Net pump current for 1, 2, and 3D at $T$ = 4.2 K as a function 
         of the driving amplitude $V_{\rm ac}$ and chemical potential $\mu$. 
		 Parameters: $\hbar\omega$ = 0.1 meV $\approx$ 24 GHz, 
		 $V_{\rm b}$ = 10, $V_{\rm qw}$ = -4, $\Delta V$ = -1 (meV), 
		 $d_{\rm b}$ = 10, $d_{\rm qw}$ = 12.5 (nm).}
\label{Current-High-T}
\end{figure}

\begin{figure}
\centerline{\resizebox{5.5cm}{!}{\includegraphics{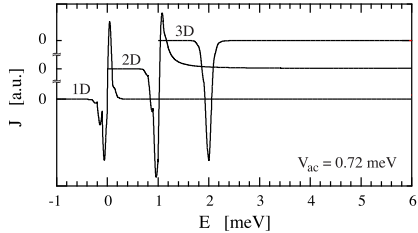}}}
\caption{Spectral current density $J(E)$ [see 
         Eq.~(\protect\ref{eq:Current})] as function of the {\it total}
		 energy of the incident electron, for 1, 2, and 3D. 
		 Curves are displaced for clarity.
		 Parameters as in Fig.~\protect\ref{Current-High-T}.}
\label{SpectralCurrentDensity}
\end{figure}

\section{Numerical results}
\label{Sec:Numerical-results}

A full numerical analysis based on transfer matrices \cite{wagner95a} fully 
confirms the results of the pipeline model: Figure~\ref{Current-High-T} 
shows the results for the pump current as a function of driving strength 
$V_{\rm ac}$ and chemical potential $\mu$ at an elevated temperature of 
$4.2\,$K and a photon energy of $\hbar\omega$ = $0.1\,$meV. The analytical 
results of the pipeline model of Fig.~\ref{Structure}, panel B, are clearly 
identified in Fig.~\ref{Current-High-T} when looking at cuts of constant 
driving amplitude $V_{\rm ac}$. This supports the initial observation 
[see Eq.~(\ref{eq:Current-3D})] that the pump current in 3D is largely 
independent of temperature, as long as $\mu - E_2$ $\gg$ $kT$.

As discussed in Sect.~\ref{Sec:Spectral-analysis}, the reason for this 
resilience against temperature can be most easily understood when considering 
the spectral current density. Let us first use the definition (\ref{Spectral}) 
for $J(E)$ based on incident channels. Figure~\ref{SpectralCurrentDensity} 
shows $J(E)$ in 1, 2, and 3D as a function of the total energy of the incident 
electron for $V_{\rm ac}$ = 0.72\,meV. As expected from the analytic 
discussion, in all cases do we find the largest contribution to the current 
stemming from states with small total kinetic energy close to the conduction 
band edge, although the details depend on the dimensionality of the pump. 
Consequently, for $\mu - E_2$ $\gg$ $\hbar\omega$ the bulk of the pump 
current is carried by states far below the Fermi surface and is insensitive 
to changes in temperature, which afflict the shape of the distribution function 
at the Fermi surface only.
	
The numerical results shown in Fig.~\ref{SpectralCurrentDensity} for a 
realistic device are explained qualitatively by Eq.~(\ref{spectral-pipeline}), 
further underlining the adequacy of the pipeline model used in \cite{Wagner99a}. 
In three-dimensions Eq.~(\ref{spectral-pipeline}) yields a square function 
localized between $E_2$ and $E_1$. The rounded peak shown in 
Fig.~\ref{SpectralCurrentDensity} for the three-dimensional case indicates 
that, in practice, one rather has a continuous distribution of pipelines.

\begin{figure}
\centerline{\resizebox{8.5cm}{!}{\includegraphics{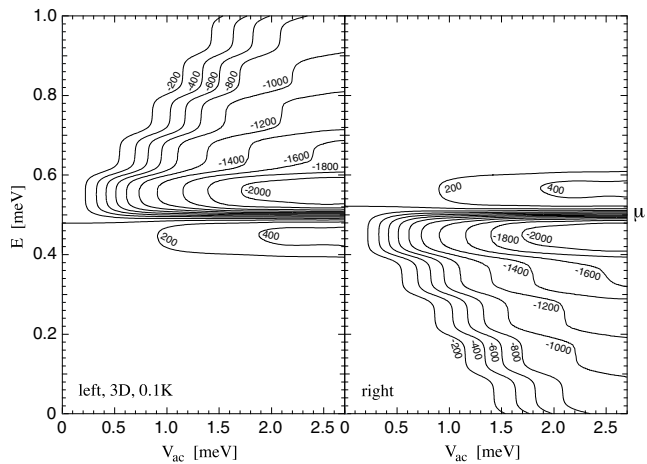}}
}
\caption{Asymptotic spectral current densities (in nA/($\mu$m$^2$ meV)) in the left 
       and right contacts of a 3D pump with $\mu$ = 0.5\,meV, $T$ = 0.1\,K,
	 $\hbar\omega$ = 0.1\,meV.
         Other parameters as in Fig.~\protect\ref{Current-High-T}. 1D and
	 2D pumps have similar characteristics at low temperatures.}
\label{AsymptoticSpectralCurrentDensity}
\end{figure}

As pointed out before, another useful definition of a spectral current 
density is based on the calculation of the net current in the leads. In this 
case we simply take the difference of the currents carried by the incident 
and outgoing channels on the same side. This is most conveniently done by 
adding all channels incoherently. We find that for
temperatures low compared to the photon energy the current density is
centred around the Fermi energy. Figure~\ref{AsymptoticSpectralCurrentDensity} 
shows the case of a 3D pump operated at the relatively low temperature
of 0.1\,K. The chemical potential was taken to be 0.5\,meV, which is somewhat
larger than the photon energy of 0.1\,meV. On the right-hand side, the
biggest contribution to the pump current stems from states about 
$\hbar\omega/2$ {\it below} the Fermi energy. These are then pumped across
the barrier and end up at energies roughly $\hbar\omega/2$ {\it above} the 
Fermi energy on the left --- indicating that the single most important
process in the pump involves a single photon, net. Very similar results are 
found for pumps of 1 and 2 spatial dimensions. This agrees well with the 
analytical expectations from Eqs.~(\ref{KLE}) and (\ref{KRE}). 
Fig.~\ref{AsymptoticSpectralCurrentDensity} also indicates that, for larger 
driving amplitudes, 
energy transfers of more than one photon quantum also contribute, although 
with a smaller weight. Finally, we note that the symmetry (\ref{symmetry}) 
around $\mu$ is very well satisfied in a realistic device, as expected from the 
analytic discussion.

\section{Conclusions}
\label{Sec:Conclusions}

We have presented a simple analytical model of electron transport through an 
asymmetric device driven by a local ac signal. The pipeline structure found 
in the electron transmission has its origin in the existence of a band offset 
between the two sides of the device. The analytic study yields an accurate 
understanding of the physics involved and shows good agreement with numerical 
results for a realistic case. A spectral analysis of the current reveals that 
the current is carried by electrons whose initial energy may be deep within 
the Fermi sea, making the pump effect resilient against temperature. On the 
other hand, the local spectral current density in the leads is localized near 
the Fermi surface. 

We appreciate helpful discussions with S.~Kohler.
This work has been supported by the EU via TMR contract
FMRX-CT98-0180, and by DGICyT (PB96-0080-C02).

\end{document}